\documentclass[letterpaper,12pt]{article}   %% LaTeX 2e (preferred)

\usepackage[colorlinks=true, citecolor=blue, linkcolor=red]{hyperref}
\usepackage{osajnl2}

\usepackage{amsmath}

\begin{document}

\newcommand{\R}[1]{\textcolor{red}{#1}}

%\twocolumn[ %% activate for two-column option

\title{A Technique for In-situ Measurement of Free Spectral Range and Transverse Mode Spacing of Optical Cavities}

%% For REVTeX it is possible to automate superscript and e-mail callouts with the superscriptaddress option; see REVTeX4 documentation.

\author{Alberto Stochino,$^{1,2,3,*}$ Koji Arai$^1$ and Rana X. Adhikari$^1$}

\address{$^1$LIGO Laboratory, MS 100-36, California Institute of Technology, Pasadena, California, 91125, USA\\
$^2$Universit\`{a} degli Studi di Siena, Dipartimento di Fisica, Via
Roma 56, 53100 Siena, Italy\\ 
$^3$ Centre for Gravitational Physics, Department of Quantum Science, The Australian National University, Canberra, Australian Capital Territory 0200, Australia\\ 
$^*$Corresponding author: stochino@ligo.caltech.edu
}

\begin{abstract} 
Length and g-factor are fundamental parameters that characterize optical cavities. We developed a technique to measure these 
parameters in-situ by determining the frequency spacing between the resonances of fundamental and spatial modes of an optical 
cavity. Two laser beams are injected into the cavity, and their relative frequency is scanned by a phase-lock loop, while the cavity 
is locked to either laser. The measurement of the amplitude of their beat note in transmission reveals the resonances of the 
longitudinal and the transverse modes of the cavity and their spacing. This method proves particularly useful to characterize 
complex optical systems, including very long and/or coupled optical cavities, as in gravitational wave interferometers. This 
technique and the results of its application to the coupled cavities of a 40-meter-long gravitational wave interferometer 
prototype are here presented.  
\end{abstract}

\ocis{120.3180, 120.2230, 120.3940, 350.1270}

%] %% activate for two-column option
The absolute length and mirror curvatures are defining parameters of an optical cavity.
Together these quantities uniquely determine the \textit{Free Spectral Range} (FSR) and
the \textit{Transverse Mode Spacing} (TMS): respectively the frequency spacing between
different resonances of the cavity's longitudinal modes and the frequency spacing between
the resonances of transverse and longitudinal modes.

FSR and TMS are usually accurately specified and it is thus very important to measure
them with great precision. Over time, monitoring cavity length and mirror curvatures can
track changes affecting the cavity geometry. For instance, variations of the cavity length
may reveal drifts of the mirrors' positions caused by ground displacement~\cite{araya99};
changes of the mirrors' curvatures can be evidence of deformations due to thermal
effects~\cite{rakhmanov04}. Precise measurements of FSR and TMS can also allow the
modeling of the cavity reflected and transmitted power by predicting the mode distribution
around the cavity working point.

Measuring these parameters proves particularly challenging in long and complex optical
cavities, such as those in gravitational wave
interferometers~\cite{ligoS6,virgostatus10,lcgtstatus10,geostatus10}. Precisely
characterizing the optical cavities in these detectors is crucial to achieve their best
sensitivity. Often three to five or more coupled optical cavities are controlled
simultaneously through the sensing of RF phase modulation sidebands added to the main laser.
Resonance or anti-resonance conditions of the sidebands in each part of the interferometer
must be ensured at all times for optimal decoupling among the degrees of freedom
\cite{redding2002dynamic}. When the RF modulation frequencies and the cavity lengths do not
match, controlling the interferometer becomes more difficult. Also, a mismatch of more
than a few centimeters in the arm cavities, or $\sim$\,1~mm in the recycling cavities may
reduce the sensitivity of the detector. Noise from the laser and the RF modulator
may then enter the interferometer's auxiliary degrees of freedom and leak into the
gravitational wave channel by cross-couplings \cite{ward2008dc,wardThesis, stochinoPhDthesis}.

Ye~\cite{Ye:04} showed that, in principle, sub-wavelength length measurement precision
could be obtained by using a femtosecond laser. However this scheme is difficult to
implement in situations where a dedicated ultra-short pulsed laser system may not bet
readily available. Several different approaches have been tried in the past. In what was
probably the simplest, Rakhmanov et al.~\cite{rakhmanov99} measured the length of a cavity
with a precision of 4~mm by an optical \textit{vernier} obtained by swinging the end
mirror. In a later experiment, Rakhmanov et al.~\cite{rakhmanov04} measured the length of a 4~km
cavity with 80~$\mu$m precision by measuring the cavity's frequency response by frequency
modulating the laser. In a similar way, Uehara and Ueda~\cite{uehara1995accurate} measured the
radius of curvature of the end mirror of a plano-concave cavity.
Additionally, Araya et al.~\cite{araya99}, following DeVoe and Brewer~\cite{devoe1984laser}, estimated
the length of a 300~meter cavity with a relative precision of $10^{-9}$ by simultaneously
locking to the cavity the laser's carrier and a phase-modulation sideband.

These techniques are difficult to scale and adapt to systems of very different lengths, or
included in complex, coupled-cavity configurations. For example, frequency modulation
techniques cannot be used in short gravitational wave interferometers since any modulation
of the laser at frequencies near the cavity FSR would be suppressed by the input mode
cleaner cavity. Tuning the sidebands' frequency as in Araya's technique, would not be
possible in systems including additional input cavities.

To circumvent these limitations, we developed an alternative interferometric technique to
measure the FSR and TMS of optical cavities with a larger range of lengths or in complex
optical setups. We then tested this technique on the optical cavities forming the LIGO
40\,m gravitational wave detector prototype at the California Institute of Technology. In
the following, we describe this technique and present the results obtained in our
experiments.

\section{Principles of the Technique}

\label{sec:fp}
The FSR of an optical cavity determines the frequency spacing $\nu_{\rm FSR}$ between resonances of any given pair of cavity longitudinal modes. It is defined as\,\cite{siegman}:
\begin{equation}
\nu_{{\rm FSR}} = \frac{c}{2L},
\end{equation}
where $L$ is the cavity length and $c$ the speed of light. This definition can then be
used to infer the cavity length from the direct measurement of the FSR. In a similar way,
the mirrors' curvatures are inferred from the measurement of the TMS and the estimated
cavity length.

Each Hermite-Gaussian mode is characterized by a different Guoy phase
determining its specific resonant frequency in the cavity
\cite{siegman}. Because of this phase, the set of resonances of a
generic TEM$_{mn}$ mode is shifted from the resonances of the
fundamental TEM$_{00}$ mode by an integer multiple of the so-called \textit{Transverse or  Spatial Mode Spacing}. For a linear cavity this is defined as
\begin{equation}\label{eq:TMS} 
  \nu_{{\rm TMS}} = \nu_{{\rm FSR}}
  \:\frac{m+n}{\pi}\cos^{-1}\sqrt{g_1g_2}. 
\end{equation} 
where $g_{1}=\left( 1-L/R_{1} \right)$ and $g_{2}=\left( 1-L/R_{2} \right)$ are the \textit{g-parameters} of the mirrors, with $R_{1}$, $R_{2}$ representing their respective radii of curvature, and $L$ the absolute length of the cavity. The product of the g-parameters $g_1g_2$ is
often referred to as the cavity \textit{g-factor}:
\begin{equation}\label{eq:gdef} 
g = g_1 g_2 
\end{equation}
It follows from (\ref{eq:TMS}) that if the cavity mirrors are astigmatic, the resonances of complimentary modes, TEM$_{mn}$ and TEM$_{nm}$ are split. If we assume, for simplicity, that axes of the astigmatism for the two mirrors are aligned, different g-factors $g_{\rm x}$ and $g_{\rm y}$ can be associated with each of the two transverse spatial directions $x$ and $y$, respectively\footnote{For more general astigmatic cavity cases, see~\cite{habraken07}.}. In case of astigmatic cavities, the definition of transverse mode spacing is then generalized as
\begin{align}
\nu_{{\rm TMS}} &= \nu_{{\rm FSR}} \left[\frac{m}{\pi}\cos^{-1}\sqrt{g_{1x}g_{2x}} + \frac{n}{\pi}\cos^{-1}\sqrt{g_{1y}g_{2y}} \right] \\
& = m\,\nu_{{\rm TMS,x}} + n\,\nu_{{\rm TMS,y}}\,\,\,, 
\end{align}
where $g_{ix} = \left(1-L/R_{ix}\right)$, $g_{iy} = \left(1-L/R_{iy}\right)$, and
$R_{ix}$, $R_{iy}$ represent the radius of curvature of the \textit{i}-th mirror of the
cavity in the $x$ and $y$ direction, respectively.

With this definition, the direct measurement of the TMS can then be used to estimate the cavity g-factor.

\subsection{Measurement technique}

The technique presented in this article determines the FSR and the TMS of a cavity from
the resonances appearing in transmission as the laser frequency is scanned. Two lasers are
used for the measurement: the first, serving as a master laser, is set to resonate in the
cavity in its TEM$_{00}$ mode; the second, phase-locked to the first, is held at an
arbitrary offset frequency set by a local oscillator (LO) in the loop~\cite{armor}. The
phase-locked loop (PLL) ensures that the relative frequency of the two lasers remains
constant.

At first, the optical cavity under test is locked to the fundamental mode of the master
laser by using the Pound-Drever-Hall technique~\cite{drever83}. The slave beam, after
being appropriately mode matched, is injected into the cavity together with the main beam.
At the cavity transmission, the two beams interfere producing a beat note at their
differential frequency as set by the PLL's LO frequency. Finally, the cavity resonance
profile is measured by tracking the beat note's amplitude as a function of the LO
frequency.

\subsection{Cavity absolute length measurement} 

As the frequency of the PLL's LO is swept, a resonance peak appears in transmission every
time the relative frequency of the slave laser reaches a multiple of the cavity FSR. The
FSR is then estimated by a least squares fit of this set of measured resonant frequencies
$\nu_n$ by the linear function $\nu_n=n\times \nu_{\rm FSR} $, where $n$ is the
resonance's order.

\subsection{Cavity g-factor measurement}\label{sec:gfactormeasurement} 

Following the FSR measurement, the TMS is measured by coupling the laser's fundamental
mode into both the TEM$_{00}$ cavity mode and the TEM$_{01}$ or TEM$_{10}$ spatial modes.
This is obtained by introducing a small misalignment between the input beam and the cavity
axis \cite[Sec. II]{anderson84}. In particular, the coupling into the cavity's TEM$_{10}$
mode is obtained by a tilt of the cavity axis in yaw; a coupling into the cavity's
TEM$_{01}$ mode is obtained by a tilt in pitch.

\begin{figure*}[tbhp]
\centering 
\includegraphics[width=.97\textwidth]{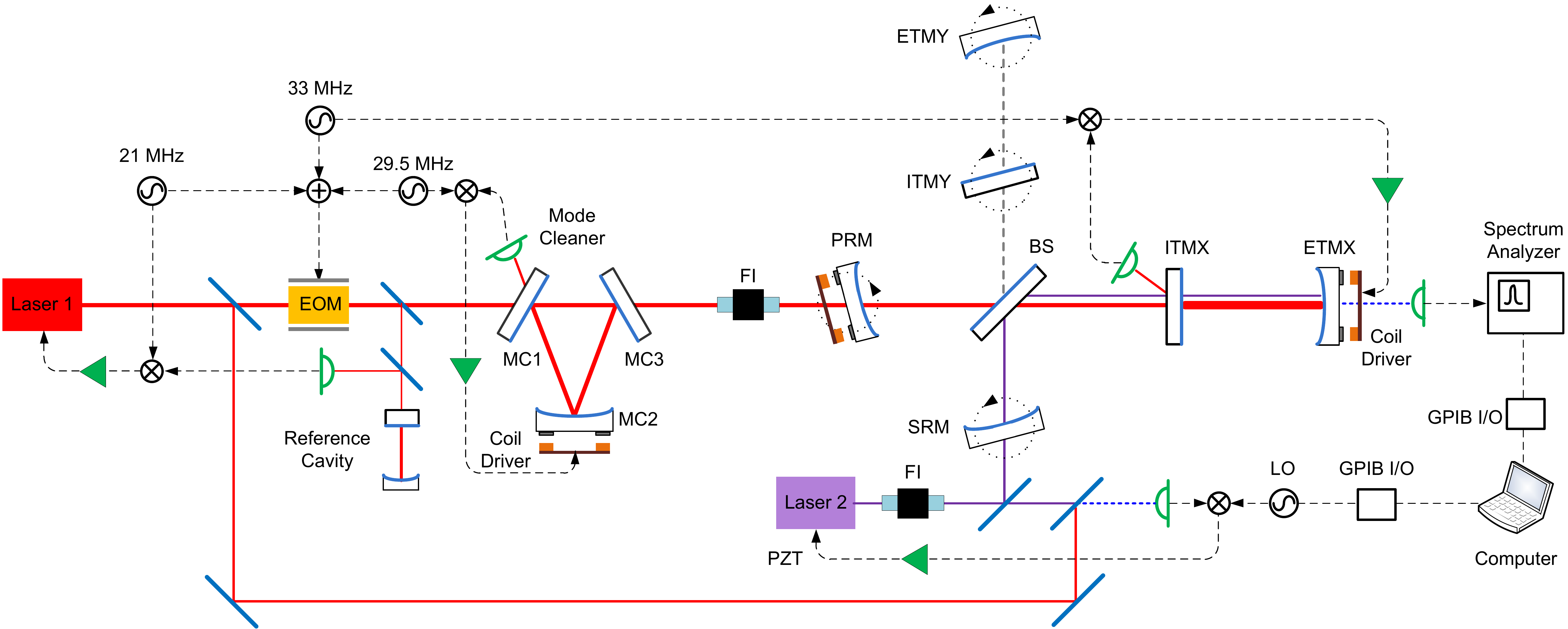}
\caption{Interferometer setup for the X arm measurement. While the arm to be measured is locked to Laser 1 (main) by PDH locking, the rest of the interferometer is held misaligned. Laser 2 is phase-locked to Laser 1 and it is then injected through the Signal Recycling Mirror (SRM). A beat note is detected in transmission by a spectrum analyzer. Its amplitude is recorded as the PLL's LO frequency is swept through several FSRs.}
\label{fig:armmeas}
\end{figure*}

The measurement begins with the relative frequency of the two lasers set to a resonant
frequency $\nu_n$ equal to an integer multiple $n$ of the cavity FSR: $\nu_n=n\times
\nu_{\rm FSR} $. Then, the frequency of the auxiliary laser is scanned, while the cavity
is held locked to the main laser. As their relative frequency equals the cavity's TMS, the
auxiliary laser resonates in the cavity's TEM$_{01}$ (or TEM$_{10}$) mode and is
transmitted by the cavity. The interference between the main beam's TEM$_{00} $ mode and
the auxiliary beam's TEM$_{01}$ (or TEM$_{10}$) mode produces a beat note on a
photodetector at a frequency $\nu=\nu_n+\nu_{\rm TMS}$. The TMS is then measured by the
frequency of the corresponding peak.

Both the TEM$_{01}$ and TEM$_{10}$ modes have antisymmetric phases around the beam axis.
As a result, their interference with the TEM$_{00}$ mode has zero net power when
integrated over the photodetector's transverse plane. To detect the beat note it is
necessary to break the symmetry of the beam spot just before the photodetector. This was
done by partially clipping the beam with a razor blade in front of the photodetector but
could also be done with a broadband quadrant photodetector.

\section{Characterization of the arm cavities of a gravitational wave interferometer} 

This technique was tested on the arm cavities of the Caltech 40\,m prototype, a testbed
facility for the Laser Interferometer Gravitational-wave Observatory
(LIGO)~\cite{weinstein2002advanced}. For the experiment, the interferometer was set in a
dual-recycled Fabry-Perot Michelson configuration. In this setup, the two ($\sim$40\,m
long) Fabry-Perot cavities, are connected in a Michelson configuration. Similar to the
Advanced LIGO interferometers~\cite{harry2010advanced}, the beam splitter's symmetric and
anti-symmetric outputs are coupled to the so-called \textit{recycling cavities}: the Power
Recycling Cavity (PRC) at the symmetric port and the Signal Recycling Cavity (SRC) at the
anti-symmetric port. All the cavity mirrors are suspended and hang on wires
as simple pendulums, for seismic noise isolation.

Figure \ref{fig:armmeas} shows the optical layout of this experiment. The main beam
illuminating the interferometer was provided by a Non-Planar-Ring-Oscillator laser (NPRO),
amplified by a Master Oscillator Power Amplifier (MOPA) up to about 2\,W at the
time of the experiment \cite{willke2008stabilized}. Dedicated frequency and power
pre-stabilization subsystems were enabled on the laser. After these stages, phase
modulation sidebands were added to the beam by electro-optic modulators (EOM) for cavity
locking. Finally, prior to entering the interferometer, the laser beam passed through a
13-meter-long triangular mode cleaning cavity. The PM sideband frequencies are chosen to
coincide with one of the FSRs of the mode cleaning cavity.

The auxiliary beam was injected into the anti-symmetric port of the interferometer,
through the signal recycling cavity's mirror (SRM). This solution allowed the beam to
bypass the input mode cleaner cavity.

The auxiliary beam was provided by a 700~mW NPRO, located outside the interferometer's
vacuum envelope, on the same optics table hosting the dark port's optical setup. A simple
flipper mirror was then used to enable the auxiliary beam to enter the interferometer at
the time of the measurement.

A phase-lock optical system was set up on the main laser table by interfering pick-off
beams from the two lasers on an RF photodiode (Thorlabs PDA255 or New Focus 1611). The PLL
ensured that the auxiliary laser's frequency followed that of the main laser with an
arbitrary tuneable offset, by suppressing the relative phase fluctuations between the two
lasers.

\section{Measurement results} 
The FSR and TMS were measured separately. In both measurements, the two laser beams were
resonated only in the cavity to be examined by misaligning the other parts of the
interferometer. The cavity was then locked to the main laser by the Pound-Drever-Hall
(PDH) technique \cite{drever83}.

A computer controlled the PLL's local oscillator via a GPIB I/O interface. A script swept
the LO frequency and monitored a spectrum analyzer connected to the PD in transmission.
The power of the PD signal at the LO frequency was recorded at each step of the frequency
scan.

% Note to editor: these two figures must be together, either next or
% on top of each other. If they get stacked vertically than the left/right
% reference in the captions should be replaced with top/bottom
\begin{figure*}[tb]
\centering
\includegraphics[width=0.47\columnwidth]{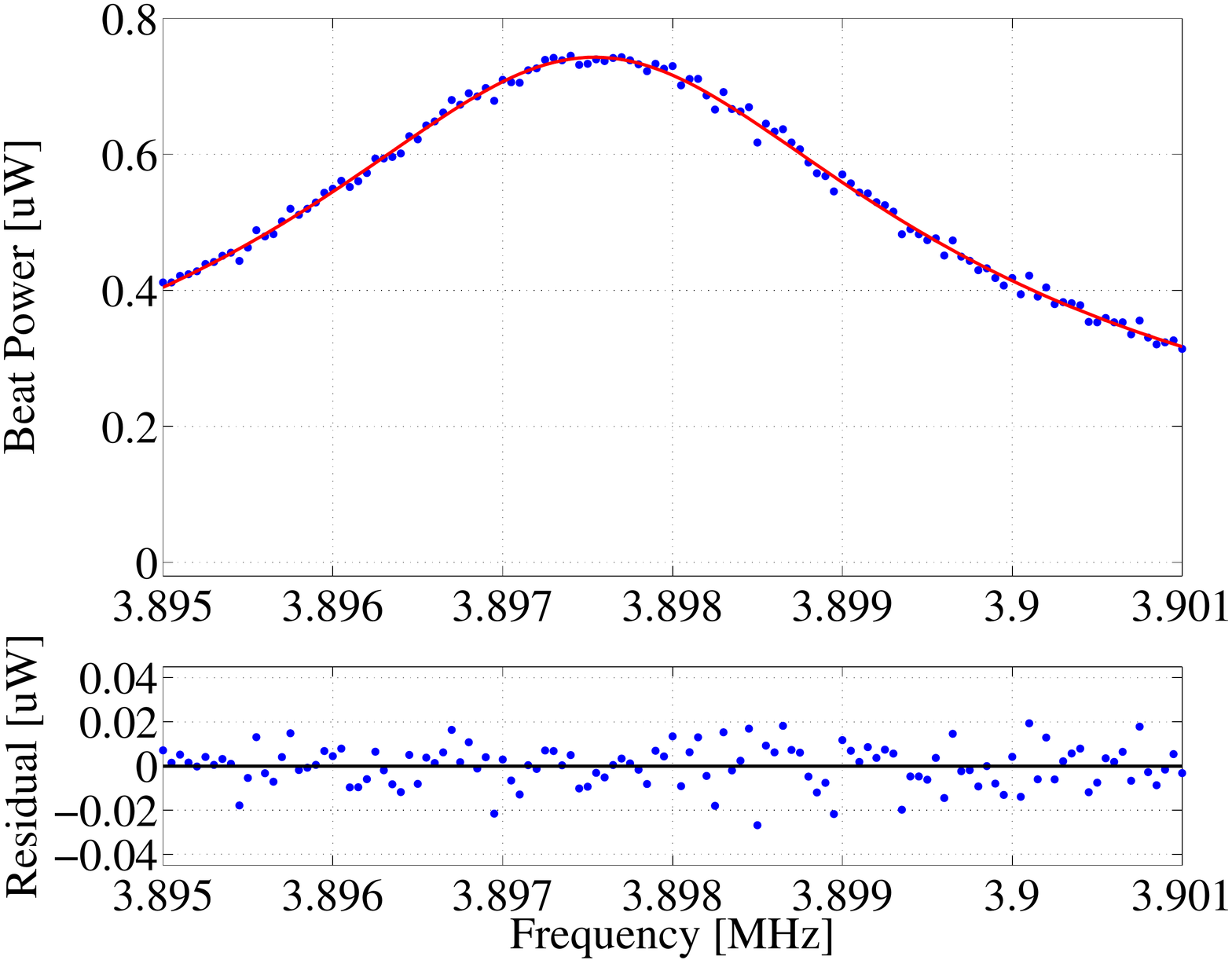}
\includegraphics[width=.5\columnwidth]{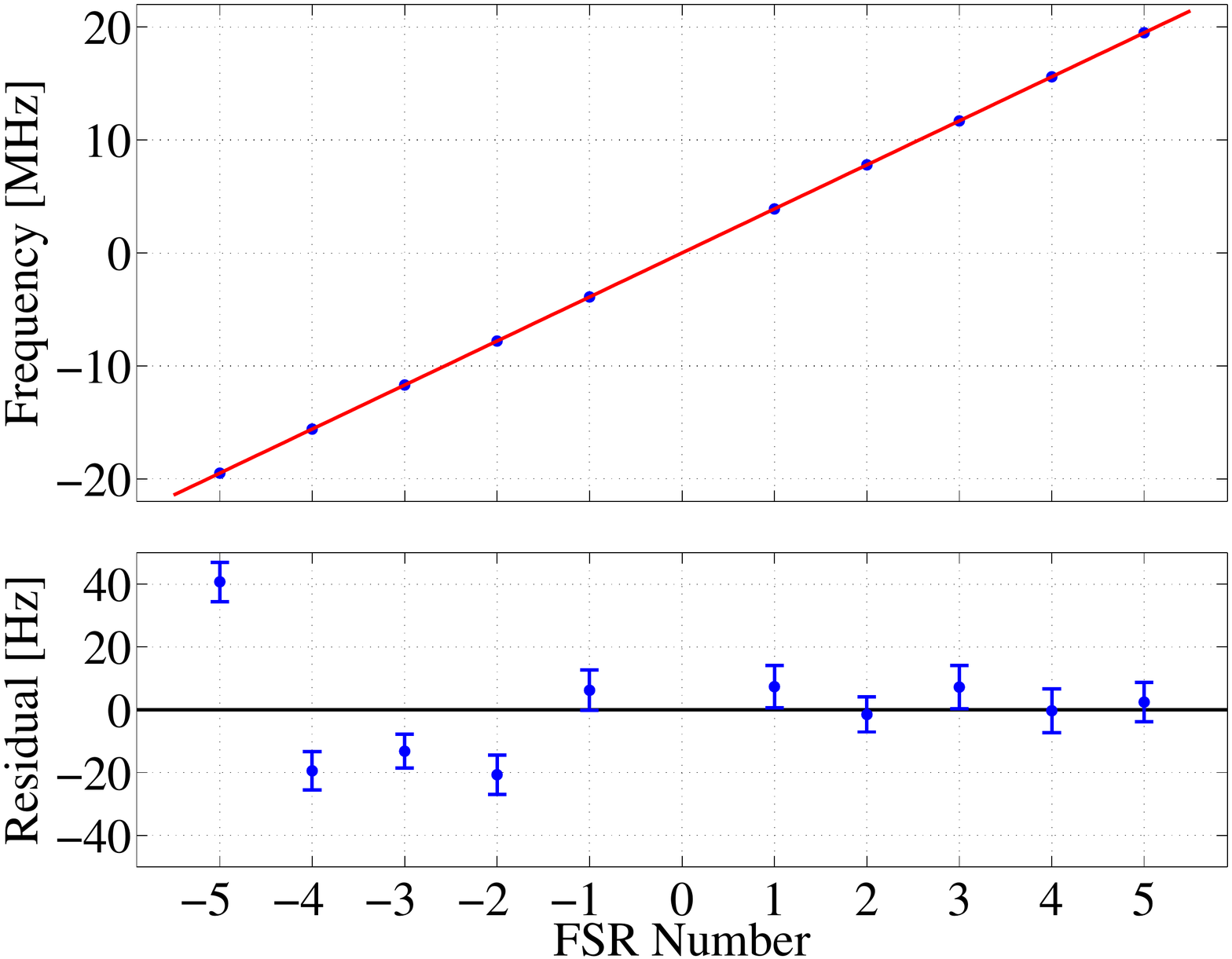}
\caption[Arm FSR scan]{Laser frequency scan of a cavity resonance
  (left) ; and linear fit of 10 resonant frequencies (right).}
\label{fig:armtransmittedpower}
\end{figure*}

\subsection{Arm length measurement} 

The arm cavity to be measured was first locked to the main laser by controlling the end
mirror with electromagnetic actuators. Then the auxiliary laser was injected into the
interferometer. The PLL frequency was scanned by $\pm$ 20~MHz, a range corresponding to
$\pm$5 cavity FSRs, by sweeping the auxiliary laser's frequency first below and then above
the main laser's frequency. The data obtained recording the power of the PD signal versus
the PLL frequency was plotted as in Fig.~\ref{fig:armtransmittedpower}. Resonance peaks
were observed at multiples of the cavity FSR. The frequency $f_0$ of each peak was
determined by a nonlinear least-squares fit of the PD signal amplitude $V_{\rm PD}$ with the function
\begin{equation}
P_{\rm PD}(f)=\frac{P_0}{\sqrt{1+\left(f-f_0\right)^2/f_c^2}}+P_{\rm off}
\label{eq:resonancemodel}
\end{equation}
where $P_0$, $P_{\rm off}$ and $f_{\rm c}$ are additional fit parameters.

By fitting 10 resonances of the X arm and 5 resonances of the Y arm, $f_0$ was estimated
with an error of $ {\sim}6-7~{\rm Hz}$ and $ {\sim}9-11~{\rm Hz}$, respectively. The
residuals from the fitting show that these errors are statistical and not systematic.

The cavity FSR was estimated by a linear least squares fit of these
resonant frequencies (right plot of
Fig.~\ref{fig:armtransmittedpower}). Table~\ref{tab:results} shows the
result of the fitting, as well as the derived values of the cavity
length. The FSRs were estimated with statistical standard errors of 0.6~Hz and 3~Hz in the X and Y arm, respectively. Accordingly, the cavity lengths were determined with a precision of $6 {\rm{\mu}m}$ and $30 {\rm{\mu}m}$, over an absolute length of $\sim$\,38.5\,m. However, the residuals in the linear fitting show deviations which are likely due to causes other than statistical fluctuations. In fact, deviations were observed at the level of $ {\sim}30-40$~Hz, equivalent to a displacement of $ {\sim}0.3-0.4$~mm (see plotted residuals in Fig. \ref{fig:armtransmittedpower}). These fluctuations may be due to drift of the cavity alignment or actual length changes during the course of the measurement due to the thermal expansion of the concrete slab.

\subsection{Arm cavity g-factor} 

The arm cavity g-factor was measured by introducing a small misalignment in the input beam's axis, first in pitch, then in yaw in order to detect the resonances of the TEM$_{01}$ mode and the TEM$_{10}$ mode, respectively. The measurement started by locking the cavity to the main laser's fundamental mode. The amplitude of the PD signal was recorded and plotted against the PLL frequency.

The frequency scan revealed additional resonance peaks due to higher order modes. The first peak was produced by the TEM$_{00}$ mode resonance; the second corresponded to the spatial mode resonance. 
\begin{figure}[tbph] 
\centering 
\includegraphics[width=1\columnwidth]{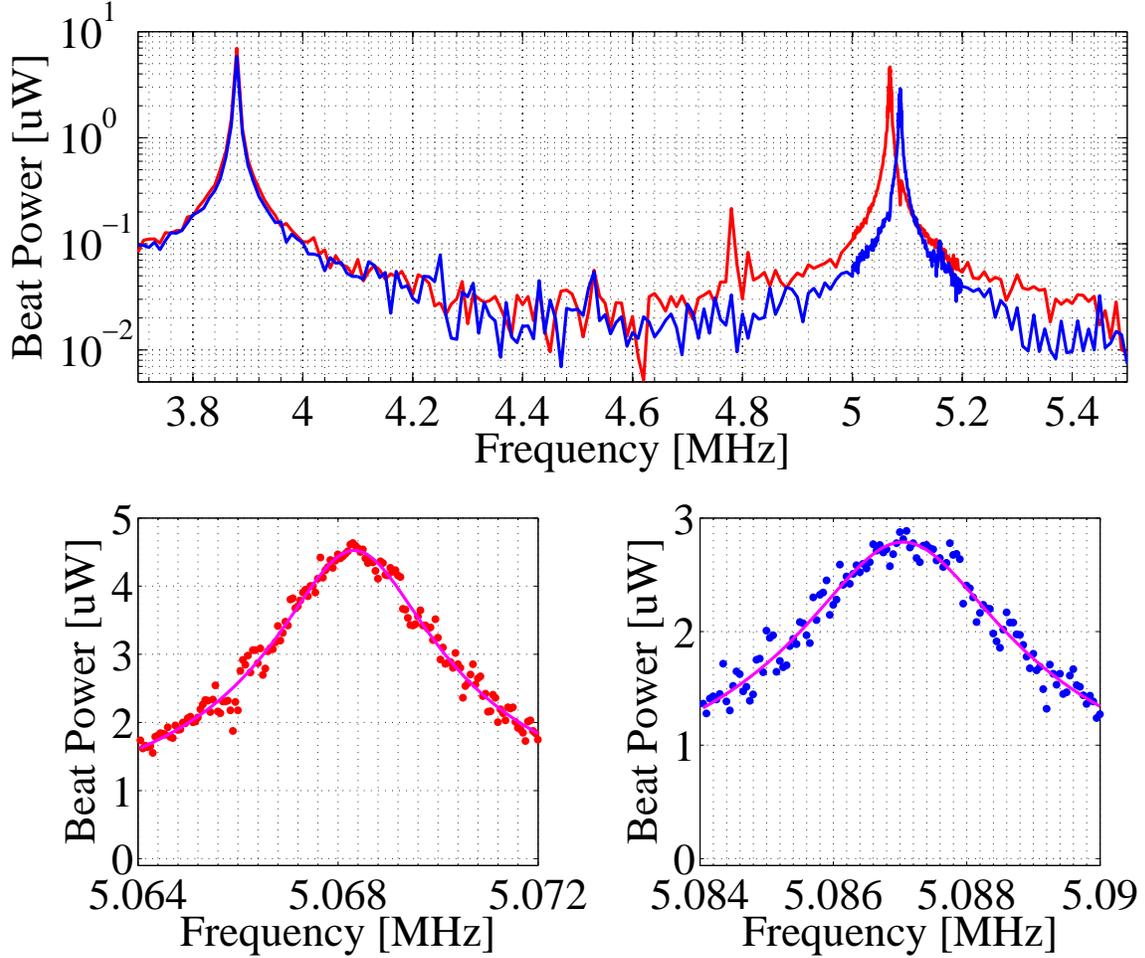} 
\caption[Measurement of transmitted TEM01 and TEM10]{Measurement of transverse mode spacing in the Y-arm. The blue and red curves are the results of the frequency scans for the horizontal and vertical modes, respectively. In the upper plot, the left peak is the cavity's fundamental resonance, while the two peaks on the right are the resonances of the spatial modes (zoomed-in in the lower plots).  A fit of the data points in these peaks (magenta lines) is used to determine their frequencies.} 
\label{fig:yarmmodes}
\end{figure}
The TMS was determined by measuring the frequencies of the TEM$_{10}$ and TEM$_{01}$ modes and then comparing them with the frequency of the TEM$_{00}$ mode.

The TEM$_{10}$ (or TEM$_{01}$) peak was identified by matching it with its expected location as calculated from the mirror's nominal radius of curvature.  The fitting of the TEM$_{10}$ and TEM$_{01}$ resonances allowed us to estimate $f_0$ with errors of 15 and 46 Hz. The fitting residuals indicate that statistical fluctuations are responsible for these deviations.

It should be noted that the g-factor measurements presented here and the values obtained by direct measurements of the mirrors' radii of curvature do not agree. Phase map measurements of the mirrors obtained by a Fizeau interferometer estimated the radii of curvature of the end and input mirrors of the X and Y arms to be 57.57, 57.68, 7280, and 7210 (all in meters), respectively. These numbers, combined with the measured cavity lengths give us g-factors of $2{\sim}6~\%$ larger than the measured values. In general, the individual radii of curvature of mirrors in a two-mirror cavity cannot be directly derived from measured g-factors. However, we can still learn something about the mirror curvatures if we take into account that typically phase map measurements of flat mirrors are more accurate than those of highly curved mirrors~\cite{Elssner:94, Elssner:89}. For instance, phase maps measurements of the nominally-flat mirrors in use in our lab have estimated radii of curvature ranging between $-100~{\rm km}$ and $+6~{\rm km}$. Finite radii like these affect the g-factors by less than $0.6~\%$, compared to an ideally flat mirror. For this reason, most likely the observed discrepancies are due to the curvature of the end mirrors. Astigmatism in the end mirrors could explain the TEM$_{01}$/TEM$_{10}$ mode splitting and justify the difference in the g-factors by $3{\sim}4~\%$. In particular, by assuming perfectly-flat input mirrors, our measurements could be explained by an astigmatism of $\sim$1~m over end mirrors with radii of curvature of 56$\sim$57~m.

The two astigmatic modes were observed in the vertical and horizontal
main axes in both arms cavities, rather than along arbitrary axes in
each of them. This might occur if the astigmatism was related to the
wedge angle on the cavity mirrors. Both the input and the output
mirrors have wedge angles of 1~deg and 2.5~deg, respectively, and they
were both installed with the angles on the horizontal axis. This
non-degeneracy of the cavity's TEM$_{10}$/TEM$_{01}$ resonances was
also confirmed by an experiment in which the cavity was locked to
either of these modes. By misaligning the cavity in pitch or yaw, it
was possible to lock the cavity in either the TEM$_{10}$ mode or the
TEM$_{01}$ but never on both at the same time. This was evidence that
the modes were indeed separated by much more than the cavity half line-width of 1.6~kHz.

\subsection{Power recycling cavity} 

Determining the length of the other cavities, such as the recycling cavity is also very important in order to achieve a stable control of the interferometer. For this reason, we tested this technique on the Power Recycling Cavity (PRC). This cavity is formed by the Power Recycling Mirror (PRM) at one end and at the other end by the Michelson ``compound mirror'', an effective mirror defined by the beam splitter and the input mirrors (ITM) of the arm cavities. The asymmetry in this short Michelson causes it to have a frequency-dependent reflectivity~\cite{Sigg:98}.

For this measurement, the auxiliary beam was introduced through a misaligned SRM. The transmitted beam was detected at the reflection port of the interferometer using the output of the Faraday isolator. The Michelson and the power recycling cavity were locked to the main laser beam, by actuating on the beam splitter and the PRM. Because the PRC length is much shorter than the arms', the laser frequency had to be scanned by several hundreds of megahertz in order to explore one cavity FSR. This was easily accomplished thanks to the large dynamic range of the PLL, in contrast to previous length measurement techniques, which had been limited to only very long cavities with smaller FSRs.

The measured profile of the transmitted power showed a complex pattern, as expected from the frequency-dependent response of the Michelson. The data were fit with an analytical model for the cavity transmissivity (details in Appendix~\ref{sec:appendix}), obtaining an estimate of the cavity absolute length and of the Michelson differential length (asymmetry) with a precision of 2~mm and 3~mm, respectively (see Table~\ref{tab:results}). 
\begin{figure}[tb] 
\centering 
\includegraphics[width=1\columnwidth]{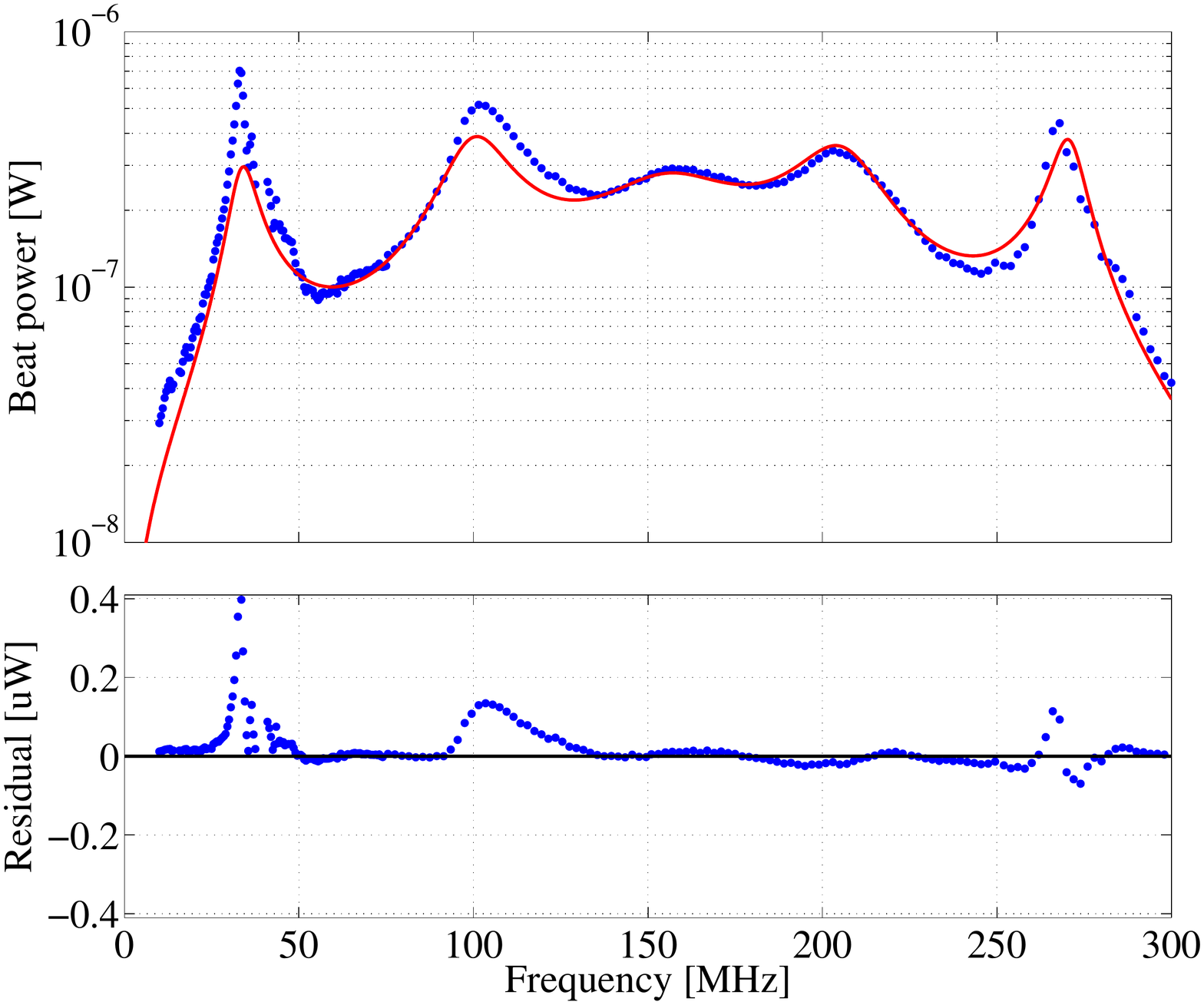} 
\caption[PRC frequency scan result]{Frequency scan of the power recycling cavity: measured beat power (blue dots) and fitting curve (red).} \label{fig:prclengthmeasresults} 
\end{figure}

\begin{table}[htbp]   
\centering   
\begin{tabular}{lrccc}  \bf{MEASURED} &&&  \\
 \hline  
\hline
\bf{X-Arm } & & & &\\
$\nu_{\rm FSR}$ & $3897627.5$& $\pm$ & $0.6$ &Hz \\
$\nu_{\rm TMS,H}$ & $1199048$\phantom{.0} & $\pm$ & $20$ &Hz\\
$\nu_{\rm TMS,V}$ & $1213602$\phantom{.0} & $\pm$ & $46$ & Hz\\
\\[-6pt]
\bf{Y-Arm}& & & &\\
$\nu_{\rm FSR}$ & $3878678$\phantom{.0}& $\pm$ & $3$ &Hz \\
$\nu_{\rm TMS,H}$ & $1207790$\phantom{.0} & $\pm$ & $23$ &Hz\\
$\nu_{\rm TMS,V}$ & $1189071$\phantom{.0} & $\pm$ & $18$ & Hz\\
\\
\bf{DERIVED}  & & &  \\  \hline  \hline
\bf{X-Arm}& & & &\\
L         &  $38.458326$ & $\pm$ & $6 \times 10^{-6}$&m \\
$g_{\rm H}$ & $0.322885$ & $\pm$ & $1.5 \times10^{-5}$\\
$g_{\rm V}$ & $0.311965$ & $\pm$ & $3.5 \times10^{-5}$\\
\\[-6pt]
\bf{Y-Arm}& & & &\\
L         &  $38.64622$ & $\pm$ & $3 \times 10^{-5}$&m \\
$g_{\rm H}$ & $0.312013$ & $\pm$ & $1.7 \times10^{-5}$\\
$g_{\rm V}$ & $0.326144$ & $\pm$ & $1.4 \times10^{-5}$\\
\\[-6pt]
\bf{PRC} & & &  \\
$l_{\rm PRC}$& $2.217$ & $\pm$ & $2 \times10^{-3} $ &m\\
Asymmetry  & $0.460$ & $\pm$ & $3 \times10^{-3} $&m \\
\hline
\end{tabular}  
\caption[Measurements summary]{Summary of measurements on the arm cavities and on the power recycling cavity. $\nu_{\rm FSR}, \nu_{\rm TMS,H},$ and $\nu_{\rm TMS,V}$ are the free spectral range and the transverse mode spacing for the horizontal mode and the vertical mode, respectively. $L$ is the cavity length. $g_{\rm H}$ and $g_{\rm V}$ are the g-factors for the horizontal and vertical modes. $l_{\rm PRC}$ and Asymmetry are the length of the power recycling cavity and the difference of the Michelson arm lengths.} 
\label{tab:results}
\end{table}   

\subsection{Measurement Precision} 

The precision of the FSR and TMS measurements listed in Table \ref{tab:results} was
determined by the statistical errors in the parameters of the
non-linear least squares fit of the
data. The difference between the two arms was due to a larger number of data points
measured for the X arm.

In principle, since the cavity is locked to the main laser, the best precision of the FSR
and TMS measurements achievable by this technique is set by the laser frequency and intensity noise and
by the noise in the PDH loop. In reality, fluctuations of the cavity stored power due to
oscillations of the cavity alignment may affect the beat note's amplitude and thus
increase the errors in the fit parameters. In our case, since the angular degrees of freedom of the
cavities were uncontrolled, the mirrors' angular motion was likely the main cause of the
measured fluctuations in the beat note signal.

In the future the technique could be greatly improved by measuring the
phase of the beat note rather than its amplitude. The cavity resonance
could then be measured with more accuracy and precision by identifying the frequency at which the
beat note's phase flips by 180 degrees. This would render
the measurement intrinsically immune to unwanted fluctuations of the
beam power. For instance the phase could be detected by
  measuring the transfer function between the PLL's local oscillator
  signal and the beat note.

\section{Conclusions} 
We demonstrated a new interferometric technique to measure the free spectral range and the
transverse mode spacing of optical cavities of various lengths. It allowed measurements of
the length and the g-factor of a Fabry-Perot cavity with a precision of 1~ppm and 10~ppm,
respectively.

Compared to previous methods, this technique can be applied to optical cavities of lengths
ranging from a few meters to several kilometers. With our experiment we proved that the
technique is compatible with a complex optical system, comprising coupled cavities. Not
requiring structural modifications to the optical system on which it is applied, it
provides a convenient option in circumstances in which an in-situ and non-invasive
measurement tool for cavity length and g-factor is needed.

\section*{Acknowledgments} 

We thank Yoichi Aso, Robert Ward and Hiro Yamamoto for illuminating discussions on this
technique. We also thank the rest of the 40\,m team for building and maintaining the
interferometer during the course of this work.

The LIGO Observatories were constructed by the California Institute of Technology and
Massachusetts Institute of Technology with funding from the National Science Foundation
under cooperative agreement PHY-0757058. This paper has been assigned a
LIGO Document Number of P1200048.

\appendix 
\section{PRC Transmissivity}
\label{sec:appendix}

When the primary laser is resonant, the transmissivity of the Power Recycling Cavity can
be written 
as a function of the PLL's local oscillator frequency 
$\Delta\omega = \omega_{\rm psl} - \omega_{\rm aux}$ in the following way:
\begin{equation}\label{eq:prcmodel}
T_{\rm prc} = \left| \frac{
t_{\rm prm}  r_{\rm itm} e^{-{\rm i} 2 \Delta\omega l_{\rm prc}/c}  \sin \left( {\Delta\omega l_{\rm asy}}/{c} \right)
}{
1 + r_{\rm prm} r_{\rm itm} e^{-{\rm i} 2 \Delta\omega l_{\rm prc}/c} \cos \left( {\Delta\omega l_{\rm asy}}/{c} \right)
} \right|^2
\end{equation}
where $t_{\rm prm}$ is the PRM transmittance; $r_{\rm itm}$ the ITM reflectance; $l_{\rm
  asy}$ the asymmetry (i.e the Michelson's differential length); and $l_{\rm prc}$ the
effective length of the power recycling cavity defined as $l_{\rm prc} \equiv l_{\rm
  prm-bs}+(l_{\rm x}+l_{\rm y})/2)$ where $l_{\rm prm-bs}$ is the distance between the
beam splitter (BS) and the PRM, and $l_{\rm x}$ and $l_{\rm y}$ are the lengths of the
Michelson arms. The amplitude of the beat note in transmission is then proportional to
$\sqrt{T_{\rm prc}}$.

\bibliographystyle{osajnl}

%\bibliography{/Users/albertostochino/mytex/mybibliography}

\end{document}